# The mechanism of solute-enriched clusters formation in neutron-irradiated pressure vessel steels: the case of Fe-Cu model alloys

A.V. Subbotin[a*] and S.V. Panyukov[b¶]

[a]Scientific and Production Complex Atomtechnoprom, Moscow 119180, Russia

[b]PN Lebedev Physics Institute, Russian Academy of Sciences, Moscow 117924, Russia

[*]e-mail: Alexey.V.Subbotin@gmail.com

[¶]Corresponding author e-mail: panyukov@lpi.ru

A B S T R A C T

Mechanism of solute-enriched clusters formation in neutron-irradiated pressure vessel steels is proposed and developed in case of Fe-Cu model alloys. We show that the obtained results are in a good agreement with available experimental data on the parameters of clusters enriched with the alloying elements. The suggested solute-drag mechanism is analogous to the well-known zone-refining process. Our model explains why the formation of solute-enriched clusters does not happen in austenitic stainless steels with fcc lattice structure. It also allows to quantify the method of evaluation of neutron irradiation dose for the process of RPV steels hardening.

Key words: solute-enriched clusters, thermal spike, solute-liquid interface, solubility, subcascade.

## 1.  Introduction

Owing to efforts of many researchers, various irradiation-induced changes in the microstructure of reactor pressure vessel (RPV) steels have been identified to date, which result in the following two main effects contributing to material embrittlement (see Refs. [1 - 5]):

- grain body hardening;
- grain boundaries weakening.

The weakening is caused by formation of radiation-induced segregations of some elements, such as phosphorus, on the boundaries of grains. Grain body hardening is caused by two factors:

- formation of ensembles of small dislocation loops and point defect clusters;
- formation of ensembles of small size (2 - 4 nm) quasi-spherical clusters enriched with solute elements, such Cu, Mn, Ni, and Si.

Objects forming the above ensembles serve to a varying degree as centers of dislocations pinning, thus causing hardening. This phenomenon raises many unavoidable questions: what are mechanisms of the objects formation? What is the strength and the lifetime of the objects as various type pinning centers?, etc. Answers to these questions should initiate solution of the problems of practical importance, such as:

- pressure vessel steels (PVS) composition optimization;
- creation of valid technique for determining neutron dose for PVS caused by the mechanisms of generation of pinning centers and formation of grain boundary segregations;
- optimization of PVS annealing modes in order to recover their properties; etc.

Ultimately, provided that the dose patterns are available both for the processes of the grain body hardening and the grain boundary weakening, the main goal of our research is to make a description of PVS embrittlement process at the polycrystalline level.

The objective of this study is to describe the mechanism of formation of solute-enriched clusters, the ensemble of which makes the main contribution to irradiation-induced hardening of RPV steels. As a result of examination [6-8] of irradiated samples of RPV steels and Fe-Cu model alloys using Atom Probe Field-Ion Microscopy (APFIM) [9], the following characteristics of clusters enriched with the solute elements were determined:

1) usually, quasi-spherical shape with the size $\bar{d} \cong 1 \div 4$ nm;

2) chemical composition of clusters formed on the basis of bcc-Fe matrix contains up to $30 \div 70$ at.% Cu, as well as Mn, Ni, and Si with concentrations exceeding by an order of magnitude the average concentration in the reference steel composition. The maximum value of the enrichment coefficient, i.e. the ratio of element concentration in a solute-enriched cluster to its reference value for copper is much higher than the enrichment coefficients of other solute elements, although their concentrations in the solute-enriched clusters can be much higher as compared to that of copper because the reference copper content may be ultimately low;

3) in contrast to precipitates, the determinant feature of the solute-enriched clusters is the absence of a pronounced concentration boundary, while copper atoms are substitution atoms for bcc-iron or bct-iron lattice (see Fig. 1);

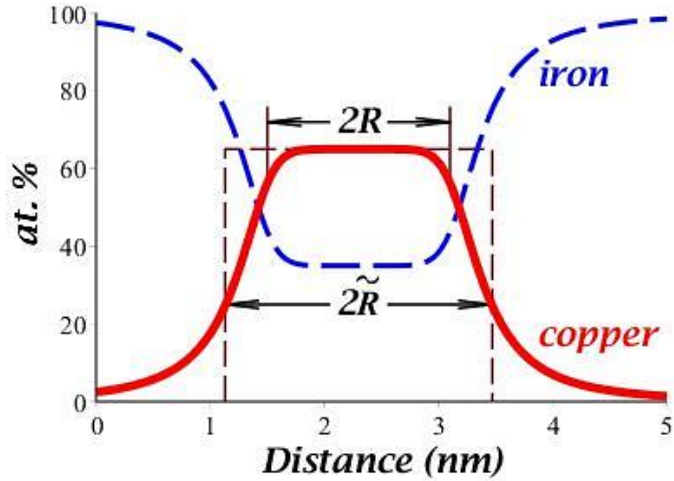

*Figure 1. Iron and copper concentration profiles for one of the clusters in neutron irradiated Fe-1.4 at.% Cu alloy [6-8], R is the "liquid phase radius" and $\tilde{R}$ is the "solute cluster radius", see Eq. (63) below.*

4) The specific temperature of PVS irradiation by neutrons that initiates formation of an ensemble of solute-enriched clusters is $T \cong 560-580$ K, at which diffusion processes in the steel are considerably suppressed. The neutron flux on PVS is also comparatively low. For the above two reasons it is rather unlikely an emergence of any collective processes, such as those occurring in case of steel irradiation in fast neutron reactors;

5) the number density of solute-enriched clusters increases with the irradiation dose reaching saturation value at the density about $10^{18}$ cm$^{-3}$.

Except vessel steels with bcc crystalline structure, bcc-Fe model alloys containing 1.4, 0.7, and 0.1 at.% Cu were irradiated and examined using APFIM in order to determine the contribution of copper to the effect under study. The observed ensembles of the copper-enriched clusters possessed all the above attributes. In this paper, the solute-enriched clusters formation model is developed by taking model alloys as examples and accounting the above considerations about low temperature and neutron flux properties. The proposed solute-enriched clusters formation mechanism is based on local high-speed processes caused by the displacements cascade due to thermal spike. Upon completion of the local processes and temperature equilibration, the formed copper-enriched cluster becomes "quasi-frozen".

## 2. The basing of the problem

In order to prove the feasibility of the proposed mechanism of solute-enriched clusters formation, the following points should be demonstrated:

- first, there is a high probability that thermal spike caused by a localized energy release in the cascade generated by the primary knock-on atom (PKA) with energy $E_{PKA} \cong 30$ keV, which has a region with average temperature significantly exceeding the melting point ($\bar{T} > T_m$) would initiate formation of a liquid phase;

- second, as the temperature decreases resulting in the supercooling of the liquid phase ($\bar{T} < T_m$), heterogeneous crystallization (liquid phase atoms joining solid-liquid interface) is more preferable process than homogeneous crystallization (viable solid phase nucleus formation).

This mechanism assures solid-liquid interface movement towards the center of the liquid phase domain and, hence, the solute-drag mechanism is implemented. In this study, the liquid phase is assumed to have quasi-spherical configuration, thus imposing no constraints on the affinity of the considered mechanism. As it was shown using molecular dynamics (MD) methods, the beginning of intensive sub-cascading is observed at PKA energies $30 \text{ keV} \leq E_{PKA} \leq 40 \text{ keV}$ [10, 11], and so $E_{PKA} \approx 30 \text{ keV}$ ($E_{MD} \approx 20 \text{ keV}$) is taken as a preliminary estimate. Nucleation of a liquid phase takes place in the superheated domain (thermal spike). The number $N_{max}$ of domain atoms and its volume $V_{max}$ linearly grow with the released energy:

$$N_{max}(3k\bar{T} + \tilde{h}) = E_{PKA}, \quad V_{max} = \frac{4}{3}\pi R_{max}^3 = \Omega N_{max} \tag{1}$$

where $\Omega$ is atomic volume, $R_{max}$ is domain radius, $3k\bar{T}$ is average thermal energy of one atom with average temperature $\bar{T}$ exceeding the Debye temperature $T_D$, $k$ is Boltzman constant, and $\tilde{h}$ is enthalpy of solid-liquid transition per atom.

We consider liquid phase nucleation process as a random walk in one-dimensional space of numbers of atoms of liquid phase particle, $n$, which are related to the particle radius $R < R_{max}$ and its volume $V$ as

$$V = \frac{4}{3}\pi R^3 = \Omega n, \tag{2}$$

Only the single-track nucleation act in $n$ space is of interest, triple and multiple collisions are ruled out as less probable. In this case, using approach described in refs. [12-16], the change of the function of atoms number distribution of phase particles $f(n,t)$ with time is described by Master equation:

$$\partial f(n,t)/\partial t = p(n-1)f(n-1,t) - q(n)f(n,t)$$
$$-p(n)f(n,t) + q(n+1)f(n+1,t) \tag{3}$$

where $p(n)$ and $q(n)$ are frequencies of, respectively, absorption ($n \to n+1$ transition) and emission ($n \to n-1$ transition) of atoms. The distribution function is normalized, $f(1,t) = N_{max}$, by the total number of atoms of the superheated domain, see Eq. (1). On condition of sufficient smoothness of functions $p(n)$ and $q(n)$, Eq. (3) can be presented as a continuity equation:

$$\frac{\partial f(n,t)}{\partial t} = -\frac{\partial I(n,t)}{\partial n},$$

$$I(n,t) = [p(n) - q(n)]f(n,t) - \frac{\partial}{\partial n}\{D(n)f(n,t)\} \tag{4}$$

where $I(n,t)$ is the nuclei flux,

$$p(n) - q(n) = \frac{dn}{dt} \tag{5}$$

is the drift component corresponding to the thermodynamically favored nucleation rate, and

$$D(n) = \frac{1}{2}[p(n) + q(n)] \tag{6}$$

is diffusion coefficient in $n$ space.

Thermal fluctuations create phase particles of different sizes. Introducing conventional equilibrium function distribution (diverging with $n \to \infty$) with Gibbs thermodynamic potential of phase particle $\Delta G(n)$,

$$f_{eq}(n) = N_{max} e^{-\Delta G(n)/kT}, \tag{7}$$

and using detailed equilibrium principle:

$$p(n)f_{eq}(n) = q(n+1)f_{eq}(n+1), \tag{8}$$

the following relation between capture and emission frequencies can be derived:

$$p(n) = q(n) \exp\left[-\frac{1}{kT}\frac{\partial \Delta G(n)}{\partial n}\right]. \tag{9}$$

Depending on which one of the two processes is identified directly on occasion, the rate of nucleation can be evaluated.

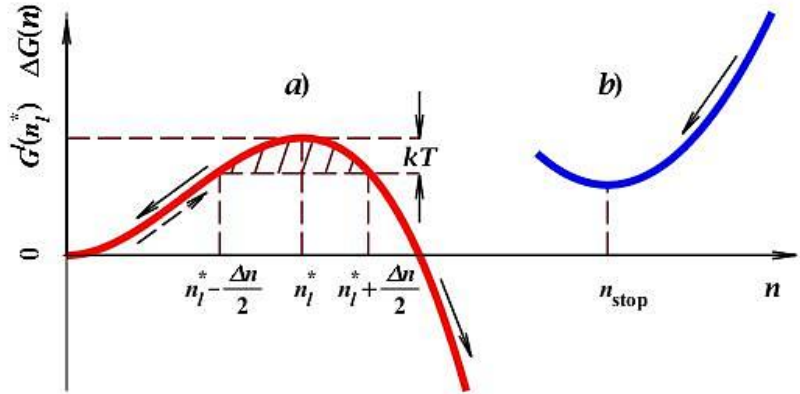

Figure 2. Thermodynamic potential of liquid phase: a) nucleation and evolution at $T > T_m$, $\Delta G^l(n_l^*) \cong 2.8kT$, $n_l^* \cong 15$ ($R_l^* \cong 3.6$ Å) for $T = 3000$ K, b) solidification at $T < T_m$, $n_{stop} \cong 300$ ($R_{stop} \cong 10.25$ Å), see section 4. Dashed arrow indicate direction of a *fluctuation process – nucleation of phase particles. Solid arrows indicate direction of the thermodynamically conditioned evolution of such particles.*

Below we consider the specific case of liquid phase nucleation ($T > T_m$). Neglecting the effect of low Cu concentration on the stage of nucleation, the following relation can be drawn for thermodynamic potential of liquid phase (see, for instance, [17] and Fig. 2):

$$\Delta G^l(n,T) = -\tilde{h}(T/T_m - 1)n + s_o \tilde{\sigma} n^{2/3}, \tag{10}$$

where $s_o \equiv 3(4\pi/3)^{2/3}$ and $\tilde{\sigma}$ is energy of solid-liquid interface per atom.

The nucleation process can be considered as a quasi steady state. Using Eqs. (3) - (10) and the presence of a sharp maximum in equilibrium distribution function $f_{eq}(n)$ at $n = n_l^*$, the following equation can be obtained for the steady state nuclei flux [12-14]:

$$I \approx \sqrt{\frac{1}{2\pi kT}\left[\frac{\partial^2 \Delta G^l(n)}{\partial n^2}\right]_{n_l^*}} D(n_l^*) f_{eq}(n_l^*), \tag{11}$$

where the number of atoms of a critical nucleus $n_l^*$ is determined from the condition that the thermodynamic potential attains a maximum:

$$\frac{\partial}{\partial n}\Delta G^l(n)\bigg|_{n_l^*} = 0, \quad n_l^* = \frac{32\pi}{3}\left[\frac{\tilde{\sigma}}{\tilde{h}(T/T_m - 1)}\right]^3 \tag{12}$$

We also have for the critical nucleus:

$$D(n_l^*) = p(n_l^*) = q(n_l^*), \quad \frac{dn}{dt}\bigg|_{n_l^*} = 0. \tag{13}$$

According to Eqs. (11) - (13), migration of phase particles with the size inside narrow interval near their critical size, $|n - n_l^*| < \Delta n/2$ (see Fig. 2 a) occurs mainly by diffusion in the space of particle sizes. The width of this interval is

$$\Delta n \approx 1\bigg/\sqrt{\frac{1}{2\pi kT}\left[\frac{\partial^2 \Delta G^l(n)}{\partial n^2}\right]_{n_l^*}}, \tag{14}$$

Further evolution of a nuclei diffused through the critical domain is thermodynamically conditioned.

In order to make numerical estimation of nucleation rate under superheating conditions ($T > T_m$), $D(n_L^*)$ term in Eq. (11) should be specified on the basis of consideration of the processes of atoms exchange between solid and liquid phases [17-20]. Substituting Eq. (2) into Eq. (5) and using Eq. (9) we find the velocity of the solid-liquid interface:

$$\frac{dR}{dt} = v_b(T)\left\{1 - \exp\left[-\frac{1}{kT}\frac{\partial \Delta G(n)}{\partial n}\right]\right\}. \tag{15}$$

In case of diffusion-controlled exchange mechanism, the velocity $v_b(T)$ is determined by Wilson-Frenkel crystal grow model:

$$v_b(T) = \frac{6a_0 D_{Fe}^l(T)}{l^2} f, \tag{16}$$

where $D_{Fe}^l(T) = \frac{kT}{3\pi\eta(t)a_0}$ is the coefficient of iron self-diffusion in liquid phase, $\eta(T)$ is the liquid phase viscosity [22, 23], $a_0$ is interatomic distance, $l$ is diffusion abrupt change length, and $f$ is the "roughness" of solid-liquid interface as a measure of capability of absorbing and emitting atoms. For the most "rough" surfaces, $f \cong 0.25$. It is rational to assume, from geometrical consideration, that this case describes the surface of quasi-spherical liquid phase, since the degree of the surface curvature is high as compared to the dimensions of atomic steps forming solid-liquid interface. Diffusion activation energy is also neglected in Eq. (16), taking into account that it is much lower than $kT$ for metals in liquid phase [17, 21, 22].

In case of significant superheating ($T > T_m$ or $T \gg T_m$), $v_b(T)$ is determined by collision-limited kinetics:

$$v_b(T) = \frac{a_0 f}{l}\sqrt{\frac{3kT}{m}}, \tag{17}$$

where $m$ is atomic mass. Studies made using molecular-dynamics (MD) simulations [23] showed significant difference between solid-liquid exchange processes under supercooling and superheating conditions. It should be noted, however, that $v_b(T)$ values calculated for superheating case using Eqs. (16) and (17) are on the same order of magnitude, and in this case, the choice of an mechanism is not crucial. In case of superheating ($T \cong 3000$ K) we estimate from Eq. (17) $v_b \cong (1.5 \div 3.0) \times 10^4$ cm/s; and in case of supercooling ($T \cong 600$ K) using $a_0 \leq l$, $\eta(T) \cong 44.4$ mPa (calculated using [12]) we obtain $v_b = (1.2 \div 2.4) \times 10^2$ cm/s.

In case of superheating using Eqs. (5), (9), and (15) we obtain:

$$\frac{dn}{dt} = q(n)\left\{\exp\left[-\frac{1}{kT}\frac{\partial \Delta G(n)}{\partial n}\right] - 1\right\}, \tag{18}$$

where $\partial \Delta G^l/\partial n > 0$ at $n < n_l^*$ and $\partial \Delta G^l/\partial n < 0$ at $n > n_l^*$. Using Eqs. (2), (13), and (15) and taking into account that $\partial \Delta G^l/\partial n < kT$, we obtain the relation between diffusion coefficient $D(n_L^*)$ and the velocity $v_b(T)$ in Eq. (17):

$$D(n_l^*) = q(n_l^*) = 3\left(\frac{4\pi}{3\Omega}\right)^{1/3}(n_l^*)^{2/3}v_b(T), \tag{19}$$

In case of solid phase nucleation in the supercooled $(T < T_m)$ liquid phase, thermodynamic potential of the solid phase is

$$\Delta G^s(n,T) = -\tilde{h}(1 - T/T_m)n + s_o\tilde{\sigma}n^{2/3}, \tag{20}$$

and similar to Eq. (19) we find:

$$D(n_s^*) \cong q(n_s^*) = 3\left(\frac{4\pi}{3\Omega}\right)^{1/3}(n_s^*)^{2/3}v_b, \tag{21}$$

where $n_s^*$ is the number of atoms in the solid phase critical nucleus, and $v_b$ is specified by Eq. (16).

Using Eqs. (11), (17) and (19), the rate of solid phase nucleation in the thermal spike (at temperature $T \cong 3000$ K) can be rewritten in the form:

$$I_0 = \frac{2V_b}{\Omega^{1/3}}\sqrt{\frac{\tilde{\sigma}}{kT}}N_{\max}\exp\left[-\frac{16\pi}{3}\left(\frac{\tilde{\sigma}}{\tilde{h}\theta}\right)^2\frac{\tilde{\sigma}}{kT}\right], \tag{22}$$

where $\theta \equiv T/T_m - 1 > 0$ and $N_{\max}$ is the number of atoms in the superheated domain, see Eq. (1). Using parameters presented in Table 1, we estimate $I_0 \cong (1.3 \div 2.6) \times 10^{15} \text{s}^{-1}$ at temperature $T = 3000$ K and $I_0 \cong (1.0 \div 2.0) \times 10^{12} \text{s}^{-1}$ at $T = 2500$ K.

| | | |
|---|---|---|
| $\Omega$ | $1.33 \times 10^{-29} \text{m}^3$ | atomic volume for solid state |
| $a$ | $2.8 \times 10^{-10}$ m | lattice parameter |
| $h$ | 13810 J·mol$^{-1}$ | enthalpy of melting |
| $\tilde{h} = \Omega h$ | 0.15 eV | enthalpy of melting per atom |
| $\sigma$ | 0.22 J·m$^2$ | solid-liquid interface energy |
| $\tilde{\sigma} = \Omega^{2/3}\sigma$ | $7.5 \times 10^{-2}$ eV | |
| $T_m$ | 1810 K | melting temperature |
| $D_T^s(T \leq T_m)$ | $(5 \div 8) \times 10^{-6} \text{m}^2\text{s}^{-1}$ | thermal diffusivity for solid |
| $D_T^l(T \geq T_m)$ | $3.2 \times 10^{-6} \text{m}^2\text{s}^{-1}$ | thermal diffusivity for liquid |
| $D_{Fe}^l/D_{Cu-Fe}^l$ | $\leq 0.3^*$ for $T \leq T_m$ | |
| $N_{\max}(\bar{T} = 3000 \text{ K})$ | $3.2 \times 10^4$ at. for $E_{PKA} \cong 30$ keV | number of superheated domain atoms |
| $R_{\max}(\bar{T} = 3000 \text{ K})$ | 47 Å for $E_{PKA} \cong 30$ keV | radius of superheated domain |
| $N_{\max}(\bar{T} = 2500 \text{ K})$ | $3.8 \times 10^4$ at. for $E_{PKA} \cong 30$ keV | |
| $R_{\max}(\bar{T} = 2500 \text{ K})$ | 49 Å for $E_{PKA} \cong 30$ keV | |

*Table 1. Iron parameters. *Estimate of the relation between iron self-diffusion in liquid phase $D_{Fe}^l$ and copper diffusion in liquid iron $D_{Cu-Fe}^l$ was taken from [22, 30, 31] by accounting the fact that activated diffusion coefficients for liquid metals are only valid as an*

*approximation for narrow temperature range $T < T_m$ near melting temperature $T_m$, where experimental data is available.*

The time of expectation of the supercritical nucleus is inversely proportional to the rate of solid phase nucleation, $\tau_n^l(T) \cong 1/I_0$, see Eq. (22). Evolution of thermal spike of different energies $E_{PKA}$ was intensively studied in refs. [24-29]. However, for our purpose it is sufficient to make only rough estimate of the thermal spike spreading time with energy $E_{PKA} \cong$ 30 keV at temperature $\bar{T} \cong 3000$ K:

$$\tau_0 \cong R_{max}^2/(2D_T^s) \cong 2 \times 10^{-12} \text{s}, \quad (23)$$

where

$$R_{max} \cong \sqrt[3]{\frac{3\Omega}{4\pi} \frac{E_{PKA}}{(3kT+\hbar)}} \cong 4.7 \times 10^{-7} \text{cm}. \quad (24)$$

Comparison of the time of viable nucleus expectation $\tau_n^l(T)$ with the thermal spike spreading time $\tau_0$ shows that $\tau_0 \gg \tau_n^l$ at temperature $T = 3000$ K, and the liquid phase nucleation takes place. At temperature $T = 2500$ K, $\tau_0 \geq \tau_n^l$, and nucleation is on the brink of its implementation.

Below we study the possibility of solid phase nucleation in the supercooled liquid phase on the stage of solidification. The lifetime of liquid phase with initial radius $R_{max}$ (see Eq. (1) and Table 1) when it is heterogeneously healed, i.e. by solid-liquid interface advancing inward the domain, can be estimated as:

$$\tau_g \cong \frac{R_{max}}{|dR/dt|}, \quad (25)$$

where the interface advancing velocity is evaluated using Eqs. (5), (8), (9), (15), and (16), with $\Delta G^l(n,T)$ defined in Eq. (10) at $T < T_m$:

$$\frac{dR}{dt} = -v_b\left[1 - \exp\left(-\frac{1}{kT}\frac{\partial \Delta G}{\partial n}\right)\right] \cong -v_b, \quad (26)$$

by taking into account that $\partial \Delta G^l/\partial n > kT$ at $T < T_m$.

Numerical solution of the thermal spike spreading problem considered as the Stefan problem showed that in spite of significant solidification enthalpy release on solid-liquid interface during the cooling-down stage, the interface movement was slow as compared to the movement of imaginary "temperature front" at $T(r(t),t) = T_m$, The liquid phase was kept substantially supercooled during the major part of $\tau_g$ time interval because the thermal diffusion coefficient $D_T^{s,l}$ was much higher than the iron diffusion coefficient in liquid phase, $D_{Fe}^l$ (see Table 1).

Using Eqs. (11), (13), (16), and (21), we may compare values of the liquid phase lifetime $\tau_g$ and the time of expectation of solid phase nucleation in liquid phase $\tau_n^s$:

$$\frac{\tau_g}{\tau_n^s} = 2 \frac{R_{max}}{\Omega^{1/3}} \sqrt{\frac{\tilde{\sigma}}{kT}} N_{max} \exp\left[-\frac{16\pi}{3}\left(\frac{\tilde{\sigma}}{\hbar\theta}\right)^2 \frac{\tilde{\sigma}}{kT}\right], \tag{27}$$

where $\theta \equiv 1 - T/T_m$. Substituting parameters at temperature $T \cong 600$ K from Table 1, this ratio can be evaluated as:

$$\tau_g/\tau_n^s \cong 0.3. \tag{28}$$

In the above estimation, the temperature $T \cong 600$ K was taken as a less favorable from the standpoint of our estimate. Therefore, the solid phase nucleation in the supercooled liquid phase during time $\tau_g$ is improbable. At higher temperatures of the liquid phase (at $T < T_m$), our estimate in Eq. (28) becomes even stronger, because temperature dependence of the equilibrium distribution function $f_{eq}(n_s^*)$ supersedes the temperature dependence of diffusion coefficient $D_{Fe}^l(T)$ in Eq. (16), which is not of activation nature. Using data from ref. [22] we can also estimate parameters of the liquid phase: viscosity of liquid iron $\eta_{Fe}(T = 600 \text{ K}) \cong 44.5$ mPa·s and diffusion coefficient of its atoms, $D_{Fe}^l(T = 600 \text{ K}) \cong 3.5 \times 10^{-10} \text{m}^2\text{s}^{-1}$.

Based on the above considerations, the following conclusions can be drawn:
- thermal spike generated by PKA with $E_{PKA} \cong 30$ keV at its initial evolution stage (on time scale $t \cong 10^{-12}$ s) can initiate nucleation of liquid phase;
- liquid phase heterogeneous solidification occurs on the supercooling stage.

## 3. The "solute drag" mechanism

Below we consider the model alloy Fe-Cu in the low-alloyed bcc-iron phase (with initial concentration $\bar{C}_{Cu} \cong 0.1$ or 1.4 at.%). As a result of cascade generation followed by thermal spike evolution, the quasi-spherical molten Fe-Cu domain is formed, which undergoes expansion and then constriction up to its complete disappearance. The end of progressive advance of the solid-liquid interface at the expansion stage is characterized by maximum molten domain extension radius $R_{max}$, see Eq. (1). Because of incommensurability of velocities of solid-liquid interface movement and Cu diffusion transfer in bcc-phase, the concentrations of Cu in liquid metal by the time of reaching the maximum radius $R_{max}$ is almost equal to that in the surrounding solid phase. Besides, on the stage of the interface backward motion, the molten solidifying domain is supercooled most of the time.

Our objective is to demonstrate that on the stage of backward motion of the solid-liquid interface, copper atoms entering the molten domain (with total number $N_{Cu} = \bar{C}_{Cu} V_{max}$) form the solute enriched cluster by means of the strong "solute drag" mechanism, with compaction of the molten domain [5, 8]. Our study is based on the following features of the solute clusters evolution in Fe-Cu model alloys:

- fairly noticeable difference between energies of solute element (copper) solubility in solid and liquid iron phases: $E^s_{sol,Cu} - E^l_{sol,Cu} \approx 0.4 \div 0.5$ eV (see below);
- mobility of the solute element in the liquid phase is much higher than that of iron ($D^l_{Fe}/D^s_{Cu-Fe} \ll 1$).

Main concepts of our analysis were previously developed in refs. [17, 32]. Under the above conditions, the copper solution in the iron liquid phase with continuously varying concentration of the solute element $C_{Cu}(t)$ due to solid-liquid interface movement, can be considered as being in quasi steady state:

$$\frac{dR}{dt} \approx -f \frac{6 a_0 D^l_{Fe}(T) C_{Fe}}{l^2} \equiv -C_{Fe} v_b(T), \tag{29}$$

where $C_{Fe} \cong 1 - C_{Cu}$ is iron concentration in liquid phase.

In order to evaluate the velocity $v_b \leq D^l_{Fe}/a_0$, we estimate the rate of $C_{Cu}$ front velocity as $v_{Cu} \cong \sqrt{D^l_{Cu-Fe}/\tau_g}$, where $D^l_{Cu-Fe}$ is coefficient of copper diffusion in liquid Fe. One can show that $v_b < v_{Cu}$ condition is met during the whole liquid phase lifetime $\tau_g \leq 10^{-10}$ s.

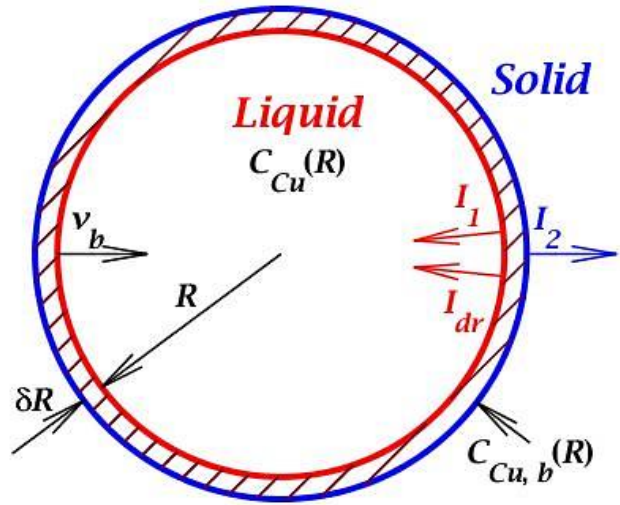

*Figure 3. Backward motion of the solid-liquid interface with velocity $v_b$: $C_{Cu}(R)$ is current Cu concentration in the liquid phase of radius R, as a result of "solute drag"; $C_{Cu,b}(R)$ is Cu concentration on the outer interface (in the solid phase); $\delta R$ is interface "thickness"; $I_1, I_2$ and $I_{dr}$ are fluxes of Cu atoms caused by interface movement.*

Interface movement leads to transfer of the solute component through it, which can be divided into three components (see Fig. 3):

1) Copper atoms flux in the direction of the interface (from the point of view of an observer on the interface). The density of this flux is:

$$I_1 = v_b C_{Fe} C_{Cu}(R). \tag{30}$$

2) The flux density of Cu atoms "leaking" through the interface:

$$I_2 = v_b C_{Fe} C_{Cu,b}(R). \tag{31}$$

3) Solute element flux due to the gradient of chemical potential $\nabla \mu_{Cu}^{s-l}(R)$ at the interface that is maintained by the interface movement. The density of such a flux, which is reflected from the interface is determined by Nernst-Einstein equation:

$$I_{dr} = -\frac{D_{Cu}^l C_{Cu}(R)}{kT} \nabla \mu_{Cu}^{s-l}(R), \tag{32}$$

We estimate the gradient in this equation as $\nabla \mu_{Cu}^{s-l} \cong (\mu_{Cu}^s - \mu_{Cu}^l)/\delta R$, where $\mu_{Cu}^s$ and $\mu_{Cu}^l$ are chemical potentials of copper in bcc-Fe and in its melt, respectively.

Chemical potential of Cu in Fe-Cu binary solution can be obtained by differentiating thermodynamic mixing potential $G$. Disregarding the mixing enthalpy in Fe-Cu case, we have:

$$G = kT(C_{Cu}\ln a_{Cu} + C_{Fe}\ln a_{Fe}), \tag{33}$$

Taking into account Gibbs-Duhem relation for the binary solution

$$C_{Cu}\frac{\partial}{\partial C_{Cu}}\ln a_{Cu} + C_{Fe}\frac{\partial}{\partial C_{Fe}}\ln a_{Fe} = 0, \tag{34}$$

and $C_{Cu} + C_{Fe} = 1$ condition, which is valid since the concentration of vacancies is small in our case, we find:

$$\mu_{Cu} = kT \ln(a_{Cu}/a_{Fe}), \tag{35}$$

where $a_{Cu}$ and $a_{Fe}$ are, respectively, Cu and Fe activities in the binary solution.

Counting chemical potential of copper atoms in liquid and solid phases from chemical potential of pure iron in bcc-phase, we find values of these chemical potentials:

$$\begin{aligned}\mu_{Cu}^s &= kT\ln(a_{Cu}^s/a_{Fe}^s) + E_{sol,Cu}^s; \\ \mu_{Cu}^l &= kT\ln(a_{Cu}^l/a_{Fe}^l) + E_{sol,Cu}^l + \tilde{h}_{Cu}(1 - T/T_m).\end{aligned} \tag{36}$$

where $\tilde{h}_{Cu}$ is enthalpy of liquid-solid state transition for Cu (see Table 1), $T_m$ is melting temperature, and $s$ and $l$ superscripts refer, respectively, to solid and liquid states. We neglect the surface term $C_{Cu}\Omega\sigma_{s-l}/R$ in $\mu_{Cu}^l$ because of smallness of $\sigma_{s-l}$, which is less than 0.22 J·m$^{-2}$ at $T < T_m$ [33].

Using Eqs. (32) and (36) we find thermodynamic barrier initiating Cu segregation process:

$$\mu_{Cu}^s - \mu_{Cu}^l \equiv \mu_{Cu}^{s-l} = kT \ln\left(\frac{a_{Cu}^s \, a_{Fe}^l}{a_{Fe}^s \, a_{Cu}^l}\right) + \delta E, \qquad (37)$$

$$\delta E = E_{sol,Cu}^s - E_{sol,Cu}^l - \tilde{h}_{Cu}(1 - T/T_m).$$

Note that segregation ("solute drag") occurs only on condition that $\mu_{Cu}^{s-l} > 0$ and at zero interface velocity an equilibrium state is reached characterized by zero chemical potential, $\mu_{Cu}^{s-l} = 0$.

Taking into account Eqs. (32) and (37), the flux of Cu atoms reflected from the interface can be rewritten in the form

$$I_{dr} = -\frac{D_{Cu-Fe}^l C_{Cu}}{\delta R} \ln\left(\frac{a_{Cu}^s \, a_{Fe}^l}{a_{Fe}^s \, a_{Cu}^l} e^{\delta E/kT}\right), \qquad (38)$$

Expression in parentheses is an analog of an equilibrium constant in absolute reaction rate theory [34]. The flux in Eq. (38), is, in general, nonlinear in $C_{Cu}$, due dependence of activity on $C_{Cu}$. From the conservation of the number of Cu atoms, we obtain (see Fig. 2):

$$I_1 + I_{dr} = I_2. \qquad (39)$$

Substituting fluxes from Eqs. (30), (31), and (38) into Eq. (39), we find the relation between Cu concentrations at both sides of the interface:

$$C_{Cu} - C_{Cu,b} \cong \frac{1}{\nu_0} \frac{D_{Cu}^l}{D_{Fe}^l} \frac{C_{Cu}}{C_{Fe}} \ln\left(\frac{a_{Cu}^s \, a_{Fe}^l}{a_{Fe}^s \, a_{Cu}^l} e^{\delta E/kT}\right), \qquad (40)$$

where $\nu_0 = 3\delta R/(8a_0)$ and interface width $\delta R \cong 3a_0$ (see [6, 7] and Fig. 3).

Taking into account that $C_{Cu,b} \leq C_{Cu,b} < 1$, and $D_{Cu}^l/(\nu_0 D_{Fe}^l) \gg 1$, one can show that the solution of Eq. (40) exists only if the expression in parentheses in Eqs. (38) and (40) (a reaction's equilibrium constant, see [34]) weakly deviated from its equilibrium value, 1:

$$\frac{a_{Cu}^s \, a_{Fe}^l}{a_{Fe}^s \, a_{Cu}^l} e^{\delta E/kT} - 1 = \varepsilon \ll 1.$$

Using this condition, Eq. (41) can be rewritten as

$$C_{Cu} - C_{Cu,b} \cong \frac{1}{\nu_0} \frac{D_{Cu}^l}{D_{Fe}^l} \frac{C_{Cu}}{C_{Fe}} \left(\frac{a_{Cu}^s \, a_{Fe}^l}{a_{Fe}^s \, a_{Cu}^l} e^{\delta E/kT} - 1\right). \qquad (41)$$

This equation relates concentrations of copper atoms in liquid phase ($C_{Cu}$) and those "remaining" in solid phase after passing the interface ($C_{Cu,b}$). We can rewrite it in a formal way

$$C_{Cu,b} = æ(Cu)C_{Cu}, \quad (42)$$

and will use below as a boundary condition on the interface to the differential equation for $C_{Cu}(R)$.

To derive this equation consider the redistribution of solute atoms due to interface motion in the domain of maximum radius $R_{max}$ of molten phase at the very beginning of the segregation process, see Eq. (1). This domain contains $N_{Cu}$ copper atoms:

$$N_{Cu} = \bar{C}_{Cu}V_{max} = \bar{C}_{Cu}(4\pi R_{max}^3/3), \quad (43)$$

where $\bar{C}_{Cu}$ is uniform initial concentration in bcc-Fe (in our case, $\bar{C}_{Cu} = 0.1$ and 1.4 at.% ). At a time $t$ during the segregation process when the radius $R(t)$ of the molten phase changes from $R_{max}$ to $\tilde{R}$, this number can be presented as

$$\int_{\tilde{V}(t)}^{V_{max}} C_{Cu,b}(t)dV + \int_0^{\tilde{V}(t)} C_{Cu}(t)dV = V_{max}\bar{C}_{Cu}. \quad (44)$$

Differentiating this equation with respect to time and keeping in mind that there is no redistribution of $C_{Cu,b}$ in solid phase due to slow diffusion at $D_{Cu}^s/D_{Cu}^l < 10^{-4}$ we find the desired differential equation for $C_{Cu}(t)$:

$$\frac{dC_{Cu}}{dt} = -\frac{C_{Cu} - C_{Cu,b}}{\tilde{V}}\frac{d\tilde{V}}{dt} \quad (45)$$

This equation is supplemented by initial conditions: $C_{Cu}(t=0) = \bar{C}_{Cu}$ and $\tilde{V}(t=0) = V_{max}$ and boundary condition (40) relating $C_{Cu}(t)$ to $C_{Cu,b}(t)$. Concentration dependence of activities $a_{Cu}^{s,l}$ and $a_{Fe}^{s,l}$ coming in Eq. (41) is shown in Table 2, see also Fig. 4.

| | | | |
|---|---|---|---|
| $A$ concentration interval | $0 \leq C_{Cu} \leq C_A$ $1 - C_A \leq C_{Fe} \leq 1$ | $a_{Cu} = \gamma_{Cu}^H C_{Cu}$ $a_{Fe} = \gamma_{Fe}^R C_{Fe}$ | Henry's law Raoultian's law |
| $B$ concentration interval | $C_A \leq C_{Cu} \leq C_B$ $1 - C_B \leq C_{Fe} \leq 1 - C_A$ | $a_{Cu} = a_{Cu}^0$ $a_{Fe} = a_{Fe}^0$ | |
| $C$ concentration interval | $C_B \leq C_{Cu} \leq 1$ $0 \leq C_{Fe} \leq 1 - C_B$ | $a_{Cu} = \gamma_{Cu}^R C_{Cu}$ $a_{Fe} = \gamma_{Fe}^H C_{Fe}$ | Raoultian's law Henry's law |
| Exemplary values | $C_A \cong 5 \div 10\ at.\%;\quad C_B \cong 95 \div 97 at.\%;$ $a_{Cu}^0 \cong 0.9 \div 0.97;\quad a_{Fe}^0 \cong 0.8 \div 0.95;$ $\gamma_{Cu}^R = \gamma_{Fe}^R = 1;\ \gamma_{Cu}^H \cong 10 \div 20;\ \gamma_{Fe}^H \cong 20 \div 35$ | | |

*Table 2 Concentration dependences of activities in Fig. 4 and exemplary values of presented parameters depending on temperature.*

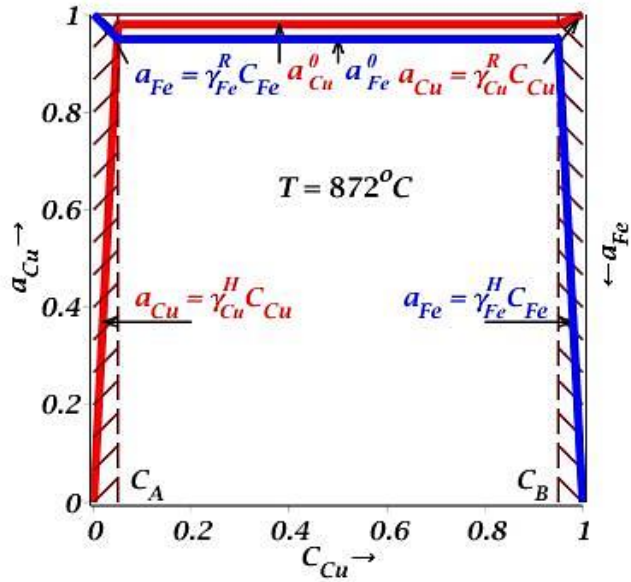

*Figure 4. Concentration dependence of activities $a_{Cu}$(from [35]) and $a_{Fe}$(calculated using [17, ch. 6] and relation $d\ln\gamma_{Fe} = -(C_{Cu}/C_{Fe})d\ln\gamma_{Cu}$).*

The solution of Eq. (45) can be divided into three different time intervals:

1) In the beginning of the solute drug process, $\bar{C}_{Cu} < C_A$ and $C_{Cu,b} \ll C_{Cu}(t)$. Using concentration dependences for $A$ interval we get $a_{Cu}^s = \gamma_{Cu,s}^H C_{Cu,b}$ and $a_{Fe}^s = \gamma_{Fe,s}^R C_{Fe,b}$ (where $C_{Fe,b} = 1 - C_{Cu,b}$ is iron residual concentration) in solid phase; $a_{Cu}^l = \gamma_{Cu,l}^H C_{Cu}$ and $a_{Fe}^l = \gamma_{Fe,l}^R C_{Fe}$ in liquid phase. Therefore, Eq. (40) takes the form:

$$C_{Cu} - C_{Cu,b} \cong \frac{1}{v_0}\frac{D_{Cu}}{D_{Fe}}\frac{C_{Cu}}{C_{Fe}}\ln\left(\frac{\gamma_{Cu,s}^H}{\gamma_{Cu,l}^H}\gamma_{Cu,s}^H e^{\delta E/kT}\frac{C_{Cu,b}C_{Fe}}{C_{Fe,b}C_{Cu}}\right). \qquad (46)$$

Using Eq. (40) and bearing in mind that $C_{Fe,b} \geq C_{Fe} \leq 1$ in A concentration interval, we find the relation between $C_{Cu}$ and $C_{Cu,b}$:

$$C_{Cu,b} = æC_{Cu}, \qquad æ \cong \left(1 + v_0 D_{Fe}^l/D_{Cu}^l\right)e^{-\delta E/kT} \qquad (47)$$

Substituting Eq. (47) into Eq. (45), we obtain the solution for A interval in the form

$$C_{Cu}(t) = \bar{C}_{Cu}[V_{\max}/\tilde{V}(t)]^{1-æ}. \qquad (48)$$

Taking into account that $C_{Cu}(t) < C_A$, the size of the molten metal domain $\tilde{R}_A$ is determined by the time at which $C_{Cu}(t) = C_A$:

$$\tilde{V}_A = V_{\max}(\bar{C}/C_A)^{1/(1-æ)}. \tag{49}$$

The residual concentration in $\tilde{R}_A \leq R \leq R_{\max}$ interval is given by

$$C_{Cu,b} = \bar{C}_{Cu}[V_{\max}/\tilde{V}(t)]^{1-æ} e^{-\delta E/kT} < C_A e^{-\delta E/kT}. \tag{50}$$

Below we give numerical estimation of main characteristics of the molten metal domain:

Minimal domain radius in this regime $\tilde{R}_A$ can be estimated using $C_A = 0.05$ (for 5 at.%) and $R_{\max} = 47$ Å. The radius $\tilde{R}_A$ at a time when $C_{Cu}(t) = C_A$ is about 13 Å - for $\bar{C}_{Cu} = 0.1$ at.% and 30 Å - for $\bar{C}_{Cu} = 1.4$ at.%.

The energy $\delta E$ (see Eq. (37)) is evaluated by taking into account that the residual concentration is $C_{Cu,b}$. Using Eq. (50) we find $\delta E \cong kT\ln(C_A/C_{Cu,b})$. Experimental data on the residual concentrations in solid phase presented in [7, 8, 36] gives $C_{Cu,b} \cong 0.03 \div 0.09$ at. %, from which it follows $0.34$ eV $\leq \delta E \leq 0.44$ eV at temperature $T \cong 1000$ K.

Using Eq. (37) for $\delta E$ and neglecting the contribution of $E^l_{sol,Cu}$, the energy of copper dissolving in bcc-iron is estimated as: $0.41$ eV $\leq E^l_{sol,Cu} \leq 0.51$ eV. This estimate is in good agreement with the experimental data [37 - 39]: $0.4$ eV $\leq E^l_{sol,Cu} \leq 0.55$ eV.

2) By the end of $C_{Cu,b} \ll C_{Cu} \leq C_A$ stage, the boundary concentration value $C_{Cu,b}$ is still in $C_{Cu} \leq C_A$ interval, whereas Cu concentration in molten metal enters into next concentration interval, $C_A \leq C_{Cu} \leq C_B$.

Using concentration dependences for A and B intervals in Eq. (40) the following equation is obtained similar to previous case 1):

$$C_{Cu} - C_{Cu,b} \cong \frac{1}{v_0}\frac{D_{Cu}}{D_{Fe}}\frac{C_{Cu}}{C_{Fe}} \ln\left(\frac{a^0_{Fe,l}}{a^0_{Cu,l}} \gamma^H_{Cu,s} e^{\delta E/kT} \frac{C_{Cu,b}}{C_{Fe,b}}\right), \tag{51}$$

and we get for the concentration range $C_A \leq C_{Cu} \leq C_B$:

$$C_{Cu,b} \cong \frac{a^0_{Cu,l}}{a^0_{Fe,l}} \frac{e^{-\delta E/kT}}{\gamma^H_{Cu,s}}. \tag{52}$$

Substituting Eq. (52) into Eq. (45), we obtain

$$C_{Cu}(t) = C_{Cu,b} + (C_A - C_{Cu,b})\left(\frac{\bar{C}}{C_A}\right)^{1/(1-æ)} \frac{V_{\max}}{\tilde{V}(t)}, \tag{53}$$

and liquid phase volume by a time of reaching the copper concentration $C_B$ is

$$\tilde{V}_B = \frac{C_A - C_{Cu,b}}{C_B - C_{Cu,b}}\left(\frac{\bar{C}}{C_A}\right)^{1/(1-æ)} V_{\max}. \tag{54}$$

Note that by this time the residual concentration $C_{Cu,b}$ in solid phase is still in A interval.

Numerical estimation for $C_A \cong 5$ at.%, $C_B \cong 97$ at.% and $R_{max} \cong 47$ Å gives $\tilde{R}_B \cong 5$ Å in case of $\bar{C} = 0.1$ at.% and $\tilde{R}_B \cong 12$ Å in case of $\bar{C} = 1.4$ at.%. The residual concentration $C_{Cu,b} \leq 0.09\%$ for the whole interval $\tilde{R}_B < R < \tilde{R}_A$.

3) By the time of reaching concentration $C_{Cu} = C_B$, the next stage starts, at which $C_{Cu} \geq C_B$ and $C_{Cu} \leq C_A$. Using concentration dependences of activities in A and C concentration intervals, we obtain in a similar way:

$$C_{Cu} - C_{Cu,b} \cong \frac{1}{v_0} \frac{D_{Cu}}{D_{Fe}} \frac{C_{Cu}}{C_{Fe}} \left[ \gamma_{Cu,s}^H \gamma_{Fe,l}^H e^{\delta E/kT} \frac{C_{Cu,b} \cdot C_{Fe}}{C_{Fe,b} \cdot C_{Cu}} - 1 \right], \quad (55)$$

and also

$$C_{Cu,b} \cong \frac{e^{-\delta E/kT}}{\gamma_{Cu,s}^H \gamma_{Fe,l}^H} \frac{C_{Cu}}{1 - C_{Cu}}, \quad (56)$$

We conclude that the residual concentration in solid phase starts to rapidly increase on this stage, which proceeds until $C_{Cu,b} \cong C_A$ condition is met. The $C_{Cu}$ value corresponding to the time of concentration $C_{Cu,b}$ entering B interval is

$$.C_{Cu}^* \cong 1 - \frac{e^{-\delta E/kT}}{C_A \gamma_{Cu,s}^H \gamma_{Fe,l}^H} \quad (57)$$

At this time $C_{Cu} \geq C_{Cu}^* \cong 0.9994$, and after that $C_{Cu,b}$ starts to rapidly increase, finally entering C interval. It certainly means the termination of the "solute drag" process. Indeed, the following relationship is fulfilled at this condition:

$$\frac{1 - C_{Cu}}{1 - C_{Cu,b}} = e^{-\delta E/kT} \quad (58)$$

Taking into account the insignificant change of the "solute cluster" volume during this stage, $C_{Cu}^*$ value can be taken as that corresponding to "solute drag" completion, and $\tilde{R}_B$ values obtained from the condition $C_{Cu} = C_B$ can be taken as "solute drag" natural stop. However, experiments show that really the "solute drag" stop occurs on essentially earlier stages. This fact can be explained by rather low reference concentrations $\bar{C}_{Cu}$ of stoppage mechanism suggested below in section 4.

Note that the volume $\tilde{V}$, within which the solid atoms segregation takes place [see Eqs. (45) - (57)], in the strict sense does not coincide with the liquid phase volume $V$ confined by the solid-liquid interface. Indeed, the $V$ domain with Cu concentration $C_{Cu}$ is surrounded by the transitional layer of thickness $\delta R$, in which the chemical potential changes with distance $0 < \zeta < \delta R$ from the solid-liquid interface as (see Eq. (37)):

$$\delta\mu_{Cu}(\zeta) \cong \mu_{Cu}(\zeta) - \mu^l = kT \ln\left[\frac{a_{Cu}(\zeta)}{a_{Fe}(\zeta)} \frac{a_{Fe}^l}{a_{Cu}^l}\right] + \delta E \frac{\zeta}{\delta R}, \tag{59}$$

So, copper concentration $C_{Cu}(\zeta)$ in this transition layer can be presented as follows:

$$C_{Cu}(\zeta) \cong C_{Cu} \frac{a_{Fe}(\zeta)}{a_{Cu}(\zeta)} \frac{a_{Cu}^l}{a_{Fe}^l} \exp\left(-\frac{\delta E}{kT} \frac{\zeta}{\delta R}\right). \tag{60}$$

Using Eqs. (46) - (57) for the whole $0 < \zeta < \delta R$ interval, except for the thin inner layers adjacent to this internal and external boundaries with Cu concentration within the $C_A < C_{Cu}(\zeta) < C_B$ range, we get:

$$C_{Cu}(\zeta) \cong C_{Cu} \exp\left(-\frac{\delta E}{kT} \frac{\zeta}{\delta R}\right). \tag{61}$$

On the concentration profiles of solute clusters presented in refs. [6-8], one can see the spreading (see Fig. 1) corresponding to a change of the thermodynamic potential inside $0 < \zeta < \delta R$ layer. Based on this experimental data, the transition layer thickness can be estimated as $\delta R \cong 3a_0$, where $a_0$ is parameter of bcc-Fe lattice.

Below we obtain the relation between "correct" liquid phase radius $R$ and "effective" radius $\tilde{R}$ corresponding to the solute cluster with uniform copper concentration $C_{Cu}$ and totally segregated component (see Fig. 1). Implicitly, $\tilde{R}$ corresponds to the radius defined in Eqs. (49), (54), etc. Using Eq. (59), we obtain:

$$\frac{4}{3}\pi \tilde{R}^3 C_{Cu} = \frac{4}{3}\pi R^3 C_{Cu} + \int_R^{R+\delta R} C_{Cu} \exp\left(-\frac{\delta E}{kT} \frac{r-R}{\delta R}\right) 4\pi r^2 dr, \tag{62}$$

Taking into account that on final stages $\delta E \gg kT$ and $\delta R/R \cong 1$, the "solute cluster" radius can be estimated as

$$\tilde{R} \cong R_s \left[1 + \left(\frac{\delta R}{R}\right)/\left(\frac{\delta E}{kT}\right)\right], \tag{63}$$

where $R_s$ is the liquid phase radius, on which the solid-liquid interface stops (see section 4).

## 4. "Solute-drag" process termination and estimates

The above considered mechanism of forced segregation termination related to an attenuation of thermodynamic barrier with increasing copper concentration, $C_{Cu}$, results in too high $C_{Cu}$ in the cluster to the end of the segregation process and too small cluster radius, in strong contradiction with available experimental data (see section 3 and [6-8]). The conclusion can be made that some other mechanisms exist, which would terminate the process somewhat earlier, i.e. at $C_{Cu} \cong 0.3 \div 0.9$ [6-8]). Bearing in mind that solute drag implementation requires two conditions to be met, namely: the presence of a thermodynamic barrier and movement of solid-liquid interface, below we consider by taking into account Eqs. (18) and (26) how the thermodynamic potential of the whole system changes in the course of evolution of the molten domain with radius $R \leq R_{\max}$. Similarly to the above consideration, we study spherical domain of radius $R_{\max}$, consisting of a melt region with radius $R < R_{\max}$ as well as peripheral solid phase with $R \leq r \leq R_{\max}$. Strictly speaking, such a system is not closed because of its non-adiabaticity and presence of elastic stress fields out of our $R_{\max}$ domain. However, on the most interesting later evolution stages, this system can be considered as isothermal, if we neglect the effect of temperature change on thermodynamic potential, considered as a function of only concentration. In this approximation the thermodynamic potential can be written as

$$G(n) = G^l(n) + G^s(n) + E_{\text{surf}}(n) + E_{\text{el}}(n, n_v), \tag{64}$$

where $n = n_{Cu} + n_{Fe}$ is the number of atoms in liquid phase, $n_{Cu} = nC_{Cu}$, see Eqs. (50) and (55),

$$G^l(n) = n_{Cu}\tilde{\mu}^l_{Cu} + n_{Fe}\tilde{\mu}^l_{Fe} \tag{65}$$

is the access of the thermodynamic potential of liquid Fe-Cu with respect to bcc-Fe state, where

$$\begin{aligned}\tilde{\mu}^l_{Cu} &= kT \ln\left(a^l_{Cu} e^{E^l_{Cu}/kT}\right), & \tilde{\mu}^l_{Fe} &= kT \ln\left(a^l_{Fe} e^{E^l_{Fe}/kT}\right), \\ E^l_{Cu} &= \tilde{h}_{Cu}(1 - T/T_m) + E^l_{\text{sol},Cu}, & E^l_{Fe} &= \tilde{h}_{Fe}(1 - T/T_m) + E^l_{\text{sol},Fe},\end{aligned} \tag{66}$$

$E_{\text{surf}}(n)$ is surface energy of solid-liquid interface, and $E_{\text{el}}(n, n_v)$ is energy of elastic stresses due to presence of molten spherical domain and $n_v$ vacancies in the peripheral solid phase.

Thermodynamic potential of solid phase is

$$G^s(n) = \int_R^{R_{\max}} \left[C_{Cu,b}(r)\tilde{\mu}^S_{Cu}(r) + C_{Fe,b}(r)\tilde{\mu}^S_{Fe}\right] 4\pi r^2 dr \tag{67}$$

where

$$\tilde{\mu}^S_{Cu}(r) = kT \ln\left[a^S_{Cu}(r) e^{E^S_{\text{sol},Cu}/kT}\right], \quad \tilde{\mu}^l_{Fe}(r) = kT \ln\left[a^S_{Fe}(r) e^{E^S_{\text{sol},Fe}/kT}\right], \tag{68}$$

and $C_{Cu,b}$ is defined in Eqs. (50) and (52).

Consider a change of the thermodynamic potential due to transition of a single atom from liquid to solid phase. In liquid phase, taking into account Gibbs-Duhem equation (34) for the binary solution, we get:

$$-\partial G^l(n)/\partial n = -\tilde{\mu}^l_{Fe} - (\tilde{\mu}^l_{Cu} - \tilde{\mu}^l_{Fe})\partial n_{Cu}/\partial n, \tag{69}$$

where $\partial n_{Cu}/\partial n$ has different forms depending on $C_{Cu}$ value intervals (see eqs. (48) and (53)). In solid phase, taking into account (as it has been done above) that diffusion processes are much slower than those under consideration, we have:

$$\partial G^s(n)/\partial n = \tilde{\mu}^s_{Fe,b} + (\tilde{\mu}^s_{Cu,b} - \tilde{\mu}^s_{Fe,b})C_{Cu,b}. \tag{70}$$

As a result, the change of thermodynamic potential of one atom due to its transition from liquid to solid phase takes the form:

$$\frac{\partial G(n)}{\partial n} = \tilde{h}_{Fe}\left(1 - \frac{T}{T_m}\right) - kT \ln \frac{a^s_{Fe,b}}{a^l_{Fe}} - E^s_{sol,Fe} + kT \frac{\partial n_{Cu}}{\partial n} \ln \frac{a^l_{Cu}}{a^l_{Fe}}$$
$$- C_{Cu,b}\left(kT \ln \frac{a^s_{Cu,b}}{a^s_{Fe,b}} + E^s_{sol,Cu} - E^s_{sol,Fe}\right) + \frac{\partial E_{surf}(n)}{\partial n} + \frac{\partial E_{el}(n,n_v)}{\partial n}. \tag{71}$$

Making analysis of concentration terms in Eq. (71) using activity values for different $C_{Cu}$ concentration intervals, it is easy to show that in the following cases:
- (AA) $C_{Cu,b} \ll C_{Cu} \leq C_A$,
- (AB) $C_{Cu,b} < C_A$, $C_A \leq C_{Cu} < C_B$,
- (AC) $C_{Cu,b} < C_A$, $C_B < C_{Cu} < C^*_{Cu}$,

the concentration dependent part of $\partial G/\partial n$, up to higher order terms, is $\frac{\partial}{\partial n}(G^l - G^s) \approx h_{Fe}(1 - T/T_m)$, and it does not make any significant contribution to the solid-liquid interface velocity.

In case (BC) $C_A \leq C_{Cu,b} < C_B$, $C_B < C_{Cu} < 1$, with

$$C^*_{Cu,b} \cong \frac{\tilde{h}_{Fe}}{E^s_{sol,Cu}}\left(1 - \frac{T}{T_m}\right), \tag{72}$$

the derivative of the concentration term of the thermodynamic potential becomes zero:

$$\frac{\partial}{\partial n}(G^l - G^s)\Big|_{C^*_{Cu,b}} = 0. \tag{73}$$

Using Eq. (60), $C_{Cu}$ value can be obtained for which Eq. (73) condition is met:

$$C^{**}_{Cu} \cong 1 - (1 - C^*_{Cu,b})e^{-\delta E/kT}, \tag{74}$$

whence it follows that $C^{**}_{Cu} \cong 0.995$ and $C^*_{Cu,b} = 0.43$.

Thus, the influence of increasing concentration of solute element in liquid phase on the velocity of the interface advancement turns out to be too weak until the later stages, which cannot be observed in experiments. The "solute drag" process should be terminated on earlier stages, i.e. somewhere in the upper half of (AB) stage: at $C_{Cu} \cong 35 \div 95$ at.% for $\bar{C} = 1.4$ at.%, and at $C_{Cu} \leq 35$ at.% for $\bar{C} = 0.1$ at.%.

Let us now consider the possibility of the interface stop under the condition $\partial G/\partial n = 0$ (see Fig. 2 b) owing to the increase in elastic stress energy on later stages of liquid phase evolution. Initially, the superheated bcc-Fe state has the volume:

$$V_{\max} = \Omega_s(T) N_{\max}, \tag{75}$$

where $N_{\max}$ is the lattice nodes number (see eq. (1)) and $\Omega_s(T)$ is atomic volume in the superheated bcc-phase. This domain becomes a liquid containing $n_v$ less number of atoms (in case of finite number $n_v > 0$ vacancies in solid state). There are, at least, two sources of $n_v$ atoms deficit. One of them, which has been studied in sufficient detail, is formation of a cascade of certain number of Frenkel pairs $I + V$ on the ballistic stage and their further evolution on the thermal spike stage [29, 40 - 47].

It was demonstrated by MD methods that upon completion of the spontaneous recombination stage, the number of stable vacancies could be described by simple empirical expression [43]:

$$n_v \cong 5.67 E_{MD}^{\alpha}, \tag{76}$$

where $E_{MD}$ (in keV) is the energy spent on the displacement cascade ($E_{MD} \cong 0.8 E_{PKA}$ for $E_{PKA} \cong 30$ keV), $\alpha = 0.779$ for $E_{MD} \leq 20$ keV and $\alpha = 1.12$ for $E_{MD} > 20$ keV. For instance, for $E_{MD} \leq 20$ keV we estimate $n_v \cong 60$, and for $E_{MD} \cong 25$ keV we find $n_v \cong 200$.

Comparison of experimental data on the formation of vacancy loops of about $1.5 \div 2$ nm size generated by cascades (cascade trace) with the results of MD modeling of cascade domain growth on later stages of thermal spike evolution led us to the conclusion that the vacancies were swept into the tapering melt domain, thus making it possible to form a vacancy loop containing $25 \div 50$ vacancies [44-47]. So, it can be concluded that the deficit of atoms $n_v > 25 \div 50$ exists in the melt domain on later stages of its evolution.

Another source of atoms deficit in the superheated domain is the process of inelastic relaxation of excess stresses caused by superheating and first-order phase transition, by emitting interstitial dislocation loops from the surface of the domain [48, 49]. So far, however, this process has not been studied enough.

Below we consider later stages, on which the temperature evolution can be neglected. The initial condition described by Eq. (78) turns into the condition of liquid phase with volume:

$$\tilde{V}_{\max} = \Omega_l(N_{\max} - n_v) = \tilde{\Omega}_l N_{\max}, \tag{77}$$

where $\Omega_l$ is atomic volume in liquid phase, and $\tilde{\Omega}_l = \Omega_l(1 - n_v/N_{max})$ is the "effective" atomic volume.

The relative "effective" change of the lattice parameter $a$ in liquid phase can be formally presented as $\tilde{a} = a(1 + \tilde{\delta})$, where

$$\tilde{\delta} = \frac{1}{3}\left(\frac{\Delta\Omega}{\Omega_s} - \frac{\Omega_l}{\Omega_s}\frac{n_v}{N_{max}}\right), \tag{78}$$

$\Delta\Omega = \Omega_l - \Omega_s$ and $a$ is the lattice parameter in bcc-phase.

The values $\Omega_l$ and $\tilde{\delta}$ correspond to free dilatation of $\tilde{V}_{max}$ volume with unstressed boundaries. Elastic stress energy due to dilatant inclusion in the solid body can be calculated in isotropic approximation, under the condition of equality of compression moduli and the absence of shear stresses in the inclusion (the case of liquid inclusion) [50, 51]:

$$E_{el} = 3\frac{1-2\nu}{1-\nu}K\tilde{\delta}^2 V_{max}, \tag{79}$$

where $K$ is the bulk modulus and $\nu$ is the Poisson modulus ratio. Substituting Eq. (78), we find the elastic energy in Eq. (64) for the thermodynamic potential:

$$E_{el}(n, n_v) = E_0 \Omega_s N_{max}\left(\frac{\Delta\Omega}{\Omega_s} - \frac{\Omega_l}{\Omega_s}\frac{n_v}{N_{max}}\right)^2. \tag{80}$$

where $E_0 = \frac{1}{3}\frac{1-2\nu}{1-\nu}K$. The first term enclosed in brackets originates from the compression stress, and it increases with $N_{max}$ increase. The second term takes into account the matter deficit: it describes the tensile stresses and increases with $N_{max}$ decrease (for fixed $n_v$ value). The stoppage of the "solute-liquid" interface motion is determined by the condition: $\partial G(n)/\partial n = 0$, and the following expression is obtained for the number of atoms of the phase particle, $n_{stop}$, if the surface energy contribution is neglected because of its insignificance:

$$n_{stop} \approx \sqrt{\frac{E_0 \Omega_s (\Omega_l/\Omega_s)^2}{\tilde{h}_{Fe}(1-T/T_m) + E_0 \Omega_s (\Delta\Omega/\Omega_s)^2}}\, n_v, \tag{81}$$

Using values of parameters taken from literature[52]: $K = 1.71 \times 10^{11}$ Pa, $T_m = 1811$ K, $\tilde{h}_{Fe} = 0.15$ eV, $\Omega_s = 1.33 \times 10^{-23}$ cm$^3$, $T \cong 1000$ K $\div$ 600 K, and $\Omega_l \cong 1.1\,\Omega_s$, we obtain:

$$n_{stop} \approx (6.4 \div 5.75)n_v(C_{Cu}), \tag{82}$$

where $n_v(C_{Cu})$ is a function of $C_{Cu}$.

Note that MD modeling of cascades showed up, at least for bcc-Fe-Cu, the tendency to the copper-vacancy clusters formation [53, 54]. Taking into account that, for the same volume of segregation domain $\tilde{V}(t)$, the concentration $C_{Cu}$ is higher in case of higher initial copper concentration $\bar{C}_{Cu}$ (see Eqs. (48) and (53)), the hypothesis can be proposed that on later evolution stages the number of vacancies in the melt domain would also be higher. Then, in bcc-Fe-Cu with higher copper content, the "solid-liquid" interface would stop at higher radius $R_{stop}$, in accordance with Eq. (82).

We can now evaluate from experimental data the average dimensions of the domain of maximum capture of solute atoms. Proceeding from the "solute drag" model developed above, time dependence of the number $N_c(t)$ of solute clusters per unit volume is determined by differential equation:

$$\frac{d}{dt}N_c(t) = K_{\text{PKA}}[1 - \bar{V}_{\max}N_c(t)], \tag{83}$$

where $K_{\text{PKA}}$ is the rate of generation of cascade-forming PKA with energies $E_b < E_{\text{PKA}} \leq 30$ keV per volume unit, $E_b$ is the melt domain formation threshold energy value, $\bar{V}_{\max}$ is average volume of melt domain at a time of reaching its maximum value. The upper limit of PKA energy $E_{PKA}$ is determined by the beginning of pronounced sub-cascading.

Note that Eq. (83) assumes, as follows from Eqs. (50) and (52) for $C_{Cu,b}$, that only first passage of the solid-liquid interface through a certain region in space is effective for the "solute drag" mechanism. Subsequent passes of the interface through the same depleted region do not drag the solute anymore.

The solution of differential Eq. (83) is:

$$N_c(t) = \frac{1}{\bar{V}_{\max}}\left(1 - e^{-t/t_{\max}}\right), \tag{84}$$
$$\bar{V}_{\max} \cong 1/N_{\max}, \quad t_{\max} \cong 1/(\bar{V}_{\max}K_{\text{PKA}}).$$

Note that in case of random arrangement of $\bar{V}_{\max}$ spheres in the volume, corresponding to the mechanism of PKA generation by the neutron flux, $\bar{V}_{\max}$ obtained from Eq. (84) gives only rather rough estimate. Using experimental value $N_{\max} \cong 2 \times 10^{18}$ cm$^3$ [1-8], we get:

$$\bar{V}_{\max} \leq 5 \times 10^{-19} \text{ cm}^3 \quad \text{and} \quad \bar{R}_{\max} \leq 50 \text{ Å}. \tag{85}$$

This result is in a good agreement with that obtained earlier (see Eq. (1)).

Below we use our advanced "solute drag" model for interpretation of available experimental data, which can be summarized as proposed below [1-4]. In neutron irradiation conditions of the model alloy containing 1.4 at.% Cu, the average copper content in the solute cloud is about 72 at.%, linear dimensions are within 2÷2.5 nm range (see also Fig. 1), the spread of concentration values between different solute clouds is in the range 45 at.% $< C_{Cu} <$ 95 at.%, and residual concentration in copper matrix is $C_{Cu,b} \cong 0.08 \div 0.02$ at.%.

As follows from this data, the solute cluster evolution is terminated at copper concentration $C_A < C_{Cu} < C_B$ and $C_{Cu,b} < C_A$ corresponding to (AB) case (see Eqs. (51) – (54)). The specific "solute cluster" presented in Fig. 1 can be described using Eqs. (53), (63), (76), and (81). The developed above description of the termination mechanism is used to find the radius $R$, and after that the "effective radius" $\tilde{R}$ (Eq. (65)) is used for evaluation of concentration $C_{Cu}$. In this case, with $\bar{C} = 1.4$ at.%, we use the following values (coordinating them with the data from Fig. 1): $R_{max} \cong 42$ Å, and $n_v \cong 60$, whence it follows that $R \cong 10.25$ Å, $\tilde{R} \cong 12$ Å, and $C_{Cu} \cong 60$ at.%.

This $R_{max}$ value corresponds to $E_{PKA} \cong 20$ keV, and $n_v$ coincides with the estimate given in Eq. (76). Note that $n_v$ is a fluctuating parameter, and $R_{max}$ is determined by $E_{PKA}$ and geometry of the thermal spike events that occurred in the irradiated medium. Variations of these two parameters would give the observed range of linear dimensions and concentrations. Model alloy containing about 0.1 at.% Cu is characterized by solute clouds with approximately the same typical dimensions (~2 nm) and copper concentration $C_{Cu} \leq 35$ at. %. The upper limit of copper concentration (35 at.%) corresponds to $R_{max} \cong 49$ Å, $\tilde{R} \cong 7$ Å, and $R \cong 5.2$ Å ($n_v \cong 8$).

As it has been mentioned above, there is correlation between copper concentrations and concentration of vacancies forming bound states with copper atoms. Probably, this is a reason of smaller $n_v$ values at lower copper concentration, and also it may result in solid-liquid interface stop at higher concentration in alloys with higher $R$ values.

## 5. Conclusions

In this paper we propose and solve a new model of "solute drag" process initiated by thermal spike in order to explain the available regular patterns obtained in experiments:
- level of copper atoms concentration in solute clusters;
- concentration pattern on the solute cluster boundary;
- linear dimensions of the solute clusters;
- residual level of solute atoms concentration outside solute cluster domain after solid-liquid interface transit;
- kinetics of solute clusters accumulation under neutron radiation.

Our studies show that the presence of thermal spike capable of forming liquid phase domain is a precondition of the "solute drag" mechanism implementation. This circumstance imposes some constraints on $E_{PKA}$ value. The lower bound ($E_{PKA}^{min}$) is determined by the requirement that the crystalline phase, which is superheated inside the thermal spike, should have dimensions (lifetime) and the temperature sufficient for nucleating molten metal domain (see section 2). The upper bound ($E_{PKA}^{max} \cong 30$ keV) is not so rigid, and is imposed proceeding from the results of MD modeling: for high energies, intensification of sub-cascading process would result in the decrease in temperatures and volumes of superheated single-connected domains. Besides, the number of thermal spikes capable of initiating the first-order phase transition also decreases, and direct relationship between the number of the primary thermal spikes and that of solute clusters disappears. Based on our advanced model, conclusion can be made that the Primary Knocked Atoms (PKA) generated as a result of neutron radiation and having energy within $E_{PKA}^{min} \leq E_{PKA} \leq E_{PKA}^{max}$ range are mainly involved in the generation of solute clusters causing embrittlement.

Our model allows to explain why the formation of clusters enriched with the alloying elements does not occur in austenitic stainless steels with fcc lattice structure. In such steels the excess free energy of the solute is small, and therefore, the solute-drag mechanism is not working.

The solute-drag mechanism discussed in this paper is analogous to the well-known zone-refining process [17] widely employed for maximum purification of many substances, from metals to organic and inorganic compounds.

Generally accepted method of determining the radiation dose of neutrons with energies $E_n > 1$ MeV or $E_n > 0.5$ MeV is based on evaluation of radiation impact on PWR vessels (see, for instance, [55]). The model developed in this paper allows to justify and quantify this method. To do this, the number of PKA and, respectively, contributions of certain energy neutron groups can be calculated in order to represent the embrittlement process as a function of radiation dose.

One more issue, which is beyond the scope of this problem but directly associated with it, should be mentioned. As applied to the problem of embrittlement (grain body hardening), the advanced model requires comprehensive information about the structure of cold solute-enriched cluster for calculation of the "strength" of solute-enriched cluster as dislocation pinning center. For instance, MD studies have demonstrated (see [47]) the possibility of implementation of various scenarios of the end structure formation.

## Acknowledgements

Authors express their acknowledgements to Dr. T.B. Ivanova, and Dr. O.V. Ivanov from P.N. Lebedev Physical Institute for fruitful discussions and assistance in work.